# A Tight Linearization Strategy for Zero-One Quadratic Programming Problems

Wajeb Gharibi[1], Yong XIA[2],

[1] *Dept. of Computer Science, College of Computer Science & Information Systems, Jazan University, Jazan 82822-6694, KSA.*
gharibi@jazanu.edu.sa

[2] *School of Mathematics and System Sciences, Beihang University*
*Beijing, 100191, P. R. China*
dearyxia@gmail.com

**Abstract**
In this paper, we present a new approach to linearizing zero-one quadratic minimization problem which has many applications in computer science and communications. Our algorithm is based on the observation that the quadratic term of zero-one variables has two equivalent piece-wise formulations, convex and concave cases. The convex piece-wise objective function and/or constraints play a great role in deducing small linearization. Further tight strategies are also discussed.

## 1. Introduction

In this article, we consider the zero-one quadratic programming problem

P:   min   $c^T x + x^T Q x$        (1.1)

s.t.   $h^T x + x^T G x \geq g$     (1.2)

$x \in X \subseteq \{0,1\}^n,$      (1.3)

where $Q$ and $G$ are general symmetric matrices of dimension $n \times n$.

This problem is a generalization of unconstrained zero-one quadratic problems, zero-one quadratic knapsack problems, quadratic assignment problems and so on. It is clearly NP-hard.

Linearization strategies are to reformulate the zero-one quadratic programs as equivalent mixed-integer programming problems (1.1) and (1.3) with additional binary variables and/or continuous variables and continuous constraints, see [1, 2, 3, 6, 7, 8, 9, 10, 12, 13]. The main contributions of this article is to present a simple approach to linearizing zero-one quadratic minimization problem. It is based on the observation that the quadratic term of zero-one variables has two equivalent piece-wise formulations, convex and concave cases. Small linearization is obtained based on reformulating the corresponding convex piece-wise objective function and/or constraints.

Recently, Sherali and Smith [14] developed small linearizations for (1.1) - (1.3), which is more general with structure. The linearization generated by our approach is smaller. More tight linearization strategies are proposed in this article for further improvement.

This article is organized as follows. In section 2, we shortly describe the existing efficient linearization approach. In section 3, we introduce our approach and represent the linearized model. Tight linearization strategies are developed in section 4. We conclude the paper in section 5.

## 2. The Existing Efficient Linearization Approach

Define

$\gamma^i_{\min/\max} = \min/\max\{Q_i x : x \in \bar{X}\}, \forall i,$    (2.1)

where $Q_i$ is the *i*-th row of $Q$, and $\bar{X}$ is any suitable relaxation of $X$ such that the problem (2.1) can be solved relatively easily. $\gamma_{\min/\max}$ be the vector with components $\gamma^i_{\min/\max}$, $i = 1,...,n$, and $\Gamma_{\min/\max} = diag(\gamma^i_{\min/\max})$.

Similarly, define

$\lambda^i_{\min/\max} = \min/\max\{G_i x : x \in \bar{X}\}, \forall i,$   (2.2)

and

$$\lambda_{\min/\max} = (\lambda^i_{\min/\max}, i=1,...,n)^T,$$
$$\Lambda_{\min/\max} = diag(\lambda^i_{\min/\max}, i=1,...,n).$$

Sherali and Smith [14] reformulated Problem $P$ as an equivalent bilinearly constrained bilinear problem by introducing $\gamma = Qx$ and $\lambda Gx$. Linearizing the terms $x_i \gamma_i$ and $x_i \lambda_i$ by $s'_i$ and $z'_i$ respectively, they obtained

BP:  min $c^T x + e^T s'$ (2.3)

s.t. $Qx = \gamma$ (2.4)

$h^T x + e^T z' \geq g$ (2.5)

$Gx = \lambda$ (2.6)

$\gamma^i_{\min} x_i \leq s'_i \leq \gamma^i_{\max} x_i, \forall i,$ (2.7)

$\gamma^i_{\min}(1-x_i) \leq (\gamma_i - s'_i) \leq \gamma^i_{\max}(1-x_i), \forall i,$ (2.8)

$\lambda^i_{\min} x_i \leq z'_i \leq \lambda^i_{\max} x_i, \forall i,$ (2.9)

$\lambda^i_{\min}(1-x_i) \leq (\lambda_i - z'_i) \leq \lambda^i_{\max}(1-x_i), \forall i,$ (2.10)

$x \in X$ (2.11)

where $e$ is a conformable vector of ones and the constrains (2.7) - (2.10) comes from multiplying

$\gamma_{\min} \leq \gamma \leq \gamma_{\max}, \lambda_{\min} \leq \lambda \leq \lambda_{\max}$ (2.12)

by $x_i$ and $(1-x_i)$.

BP (2.3) - (2.11) has the following equivalent compact formulation

BP:  min $c^T x + e^T s + \gamma^T_{\min} x$ (2.13)

s.t. $Qx = y + s + \Gamma_{min} e$ (2.14)

$h^T x + e^T z + \lambda^T_{\min} x \geq g$ (2.15)

$Gx = \lambda$ (2.16)

$0 \leq s_i \leq (\gamma^i_{\max} - \gamma^i_{\min}) x_i, \forall i,$ (2.17)

$0 \leq y_i \leq (\gamma^i_{\max} - \gamma^i_{\min})(1-x_i), \forall i,$ (2.18)

$0 \leq z_i \leq (\lambda^i_{\max} - \lambda^i_{\min}) x_i$ (2.19)

$\lambda^i_{\min} \leq (\lambda_i - z_i) \leq \lambda^i_{\max} - (\lambda^i_{\max} - \lambda^i_{\min}) x_i, \forall i,$ (2.20)

$x \in X$ (2.21)

via the linear transformation

$s_i = s'_i - \gamma^i_{\min} x_i, \forall i,$
$y_i = \gamma_i - s'_i - \gamma^i_{\min}(1-x_i), \forall i,$ (2.22)
$z_i = z'_i - \lambda^i_{\min} x_i, \forall i,$

Since the optimization and constraint senses of BP tend to push the variables s to their lower bounds and z to their upper bounds, the final relaxed version of BP was written as

$\overline{BP}$:  min $c^T x + e^T s + \gamma^T_{\min} x$ (2.23)

s.t. $Qx = y + s + \Gamma_{min} e$ (2.24)

$0 \leq y \leq [\Gamma_{\max} - \Gamma_{\min}](e-x)$ (2.25)

$s \geq 0$ (2.26)

$h^T x + e^T z + \lambda^T_{\min} x \geq g$ (2.27)

$Gx \geq z + \lambda_{\min}$ (2.28)

$0 \leq z \leq [\Lambda_{\max} - \Lambda_{\min}] x$ (2.29)

$x \in X,$ (2.30)

by deleting the upper bounding inequalities for $s$ and $\lambda - z$ in (2.17) and (2.20), and combining (2.16) with (2.20).

It was shown in [14] that Problems $BP$ and $P$ are equivalent in the sense that for each feasible solution to one problem, there exists a feasible solution to the other problem having the same objective value. Furthermore, let $x$ be part of an optimal solution to Problem $BP$. Then $x$ solves Problem $P$.

Besides, $BP$ can be improved by the additional cuts

$(\lambda^i_{\min} - \gamma^i_{\min} - w^i_{\max}) x_i - s_i + z_i \leq 0, \forall i,$ (2.31)

which is derived from multiplying $\lambda_i - \gamma_i \leq w^i_{\max}$ by $x_i$ where $w^i_{\max} = \max\{(G_i - Q_i)x : x \in \overline{X}\}$.

## 3. A Representation Approach

Motivated by [15], we first reveal the relation between general quadratic and piece-wise linear terms for zero-one variables.

Lemma 3.1

let $x \in X \subseteq \{0,1\}^n$. for all $i = 1,...,n$,

$x_i Q_i x = \max\{\gamma^i_{\min} x_i, Q_i x + \gamma^i_{\max} x_i - \gamma^i_{\max}\},$ (3.1)

$x_i Q_i x = \min\{\gamma^i_{\max} x_i, Q_i x + \gamma^i_{\min} x_i - \gamma^i_{\min}\}$ (3.2)

*Proof.* Suppose $x_i = 0$, the left hand side of (3.1) is clearly 0 and the right hand side becomes $\max\{0, Q_i x - \gamma^i_{\max}\} = 0$. On the other hand, if $x_i \neq 0$, it must hold that $x_i = 1$, the right hand side of (3.1) reads $\max\{\gamma^i_{\max}, Q_i x\} = Q_i x$, which is equal to the left hand side. The proof of (3.1) is completed and (3.2) can be similarly verified.

Corollary 3.1

Let $x \in X \subseteq \{0,1\}^n$. for all $i = 1,...,n$,

$$\max\{\gamma_{\min}^i x_i, Q_i x + \gamma_{\max}^i x_i - \gamma_{\max}^i\} \leq s_i' \leq \quad (3.3)$$
$$\leq \min\{\gamma_{\max}^i x_i, Q_i x + \gamma_{\min}^i x_i - \gamma_{\min}^i\},$$

if and only if

$$s_i' = x_i Q_i x. \quad (3.4)$$

*Proof.* Combining (3.1) with (3.2), we have

$$x_i Q_i x = \max\{\gamma_{\min}^i x_i, Q_i x + \gamma_{\max}^i x_i - \gamma_{\max}^i\} \quad (3.5)$$
$$= \min\{\gamma_{\max}^i x_i, Q_i x + \gamma_{\min}^i x_i - \gamma_{\min}^i\},$$

□

The above results hold true for $G_i$ and $\lambda$, defined before. Linearization based on Corollary 3.1 is just BP (2.3) - (2.11), where the linear inequalities (2.7) - (2.8) is nothing but (3.3). We remark here the four inequalities implied by (3.3) were first introduced in [8].

Actually, not all inequalities (3.3) are necessary in the final linearized model. To see this, below we first introduce the principle of reformulating zero-one quadratic programs into piece-wise linear programs. Generally, for continuous programs, we have

**Proposition 3.1**. Any convex program with linear or piece-wise linear objective function and constraints is equivalent to a linear program in the sense that there is a one-to-one projection between both feasible solutions.

*Proof.* We notice that

$$\min f(x)$$

is equivalent to

$$\min \quad t$$
$$s.t. \quad t - f(x) \geq 0$$

Without loss of generality we assume that the objective function is linear. The constraint set is convex and characterized by piece-wise linear inequalities. It follows that it is convex polyhedral, which must have linear expression. □

It is easy to see that the equivalence of Proposition 3.1 holds if we restrict the variables to be zeros or ones. Next we show the existence of such equivalent 'convex' piece-wise linear program for zero-one quadratic minimization problem.

**Proposition 3.2.** For any zero-one quadratic minimization problem, there is an equivalent zero-one piece-wise linear program with convex objective function and constraints.

Proof. Clearly, the maximum of several linear functions is convex and the minimum is concave. Then (3.1) and (3.2) in Lemma 3.1 provide the convex and concave formulations, respectively. Therefore, for any given zero-one quadratic minimization problem, we can obtain an equivalent convex piece-wise linear program by using (3.1) and/or (3.2). Note that we use (3.1) and (3.2) simultaneously only when handling equality constraints, see also Corollary 3.1.

Now we can see that (1.1) - (1.3) has the following equivalent formulation

$$\min \quad c^T x + \sum_{i=1}^n \max\{\gamma_{\min}^i x_i, Q_i x + \gamma_{\max}^i x_i - \gamma_{\max}^i\} \quad (3.6)$$

$$s.t. \; h^T x + \sum_{i=1}^n \min\{\lambda_{\max}^i x_i, G_i x + \lambda_{\min}^i x_i - \lambda_{\min}^i\} \geq g, \quad (3.7)$$

$$x \in X \subseteq \{0,1\}^n. \quad (3.8)$$

Linearizing (3.6)-(3.8) becomes very easy. For example, (3.7) is equivalent to

$$h^T x + \sum_{i=1}^n z_i \geq g, \quad (3.9)$$

$$z_i \leq \lambda_{\max}^i x_i, \quad (3.10)$$

$$z_i \leq G_i x + \lambda_{\min}^i x_i - \lambda_{\min}^i, \quad (3.11)$$

since (3.9)-(3.11) is a relaxation of (3.7) and (3.9)-(3.11) also implies (3.7).

Now we can obtain a linearization for (3.6)-(3.8), which is similarly to $\overline{BP}$ except that we do not require $y \geq 0$ and $z \geq 0$. In other words, they are redundant in $\overline{BP}$.

Finally, we point out that the non-necessity of inequalities such as $y \geq 0$ and $z \geq 0$ was also observed in [1, 2]. Actually, the linearization generated by our convex piece-wise approach coincides theirs.

## 4. Tight Strategies

We show Lemma 3.1 can be strengthened. Define

$$\overline{\gamma}_1^i = \max\{Q_i x : x \in \overline{X}, x_i = 0\}, \forall i, \quad (4.1)$$

$$\underline{\gamma}_1^i = \min\{Q_i x : x \in \overline{X}, x_i = 1\}, \forall i, \quad (4.2)$$

$$\overline{\gamma}_2^i = \max\{Q_i x : x \in \overline{X}, x_i = 1\}, \forall i, \quad (4.3)$$

$$\underline{\gamma}_2^i = \min\{Q_i x : x \in \overline{X}, x_i = 0\}, \forall i, \quad (4.4)$$

Lemma 4.1

Let $x \in X \subseteq \{0,1\}^n$. *For all* $i = 1,...,n$,

$$x_i Q_i x = \max\{\underline{\gamma}_1^i x_i, Q_i x + \overline{\gamma}_1^i x_i - \overline{\gamma}_1^i\}, \quad (4.5)$$

$$x_i Q_i x = \min\{\overline{\gamma}_2^i x_i, Q_i x + \underline{\gamma}_2^i x_i - \underline{\gamma}_2^i\}. \quad (4.6)$$

*Proof.* Suppose $x_i = 0$, the left hand side of (4.5) is clearly 0 and the right hand side becomes $\max\{0, Q_i x - \overline{\gamma}_1^i\} = 0$ due to the definition (4.1). On the other hand, if $x_i \neq 0$, it must hold that $x_i = 1$, according to (4.2), the right hand side of (4.5) reads $\max\{\underline{\gamma}_1^i, Q_i x\} = Q_i x$ which is equal to the left hand side. The proof of (4.5) is completed and (4.6) can be similarly verified. □

Corollary 4.1

Let $x \in X \subseteq \{0,1\}^n$. For all $i = 1, \ldots, n$,

$$\max\{\underline{\gamma}_1^i x_i, Q_i x + \overline{\gamma}_1^i x_i - \overline{\gamma}_1^i\} \leq s_i' \leq \quad (4.7)$$
$$\leq \min\{\overline{\gamma}_2^i x_i, Q_i x + \underline{\gamma}_2^i x_i - \underline{\gamma}_2^i\},$$

*if and only if*

$$s_i' = x_i Q_i x. \quad (4.8)$$

Similarly, define $\overline{\lambda}_1, \overline{\lambda}_2, \underline{\lambda}_1, \underline{\lambda}_2$ as follows

$$\overline{\lambda}_1^i = \max\{G_i x : x \in \overline{X}, x_i = 0\}, \forall i, \quad (4.9)$$

$$\underline{\lambda}_1^i = \min\{G_i x : x \in \overline{X}, x_i = 1\}, \forall i, \quad (4.10)$$

$$\overline{\lambda}_2^i = \max\{G_i x : x \in \overline{X}, x_i = 1\}, \forall i, \quad (4.11)$$

$$\underline{\lambda}_2^i = \min\{G_i x : x \in \overline{X}, x_i = 0\}, \forall i, \quad (4.12)$$

And let

$$\overline{\Gamma}_{1/2} = diag(\overline{\gamma}_{1/2}^i), \overline{\Lambda}_{1/2} = diag(\overline{\lambda}_{1/2}^i),$$
$$\underline{\Gamma}_{1/2} = diag(\underline{\gamma}_{1/2}^i), \underline{\Lambda}_{1/2} = diag(\underline{\lambda}_{1/2}^i).$$

As pointed by one referee, the enhancing lower and upper bounding parameters approach was first developed in [3].

Here we notice that $\overline{\gamma}_1^i, \underline{\gamma}_1^i, \overline{\gamma}_2^i, \underline{\gamma}_2^i$ can be further enhanced. We replace $\overline{X}$ in the linear problems (4.1) - (4.4) with a more restricted feasible region:

$$\overline{X} \cap conv\{x : h^T x + x^T G x \geq g\} \quad (4.13)$$

$$\subseteq \overline{X} \cap conv\{x : h^T x + \sum_{i=1}^n \min\{\overline{\gamma}_2^i x_i, G_i x + \underline{\gamma}_2^i x_i - \underline{\gamma}_2^i\} \geq g\} \quad (4.14)$$

$$= \{x \in \overline{X} : h^T x + \quad (4.15)$$
$$+ \sum_{i=1}^n y_i \geq g, y_i \leq \overline{\gamma}_2^i x_i, y_i \leq G_i x + \underline{\gamma}_2^i x_i - \underline{\gamma}_2^i, \forall i\},$$

where $conv\{S\}$ denotes the convex hull of the set $S$, and (4.13) - (4.14) holds due to the fact

**Proposition 4.1.** The concave envelope of the bilinear function $x_i G_i x$ over the domain $[0,1] \times [\underline{\gamma}_2^i, \overline{\gamma}_2^i]$ is given by

$$\min\{\overline{\gamma}_2^i x_i, G_i x + \underline{\gamma}_2^i x_i - \underline{\gamma}_2^i\} \quad (4.16)$$

The proof of Proposition 4.1 is omitted here. General results for bilinear function and proofs could be found in [5, 11].

Similarly, (1.1)-(1.3) is equivalent to

$$\min \quad c^T x + \sum_{i=1}^n \max\{\underline{\gamma}_1^i x_i, Q_i x + \overline{\gamma}_1^i x_i - \overline{\gamma}_1^i\} \quad (4.17)$$

$$s.t. \quad h^T x + \sum_{i=1}^n \min\{\overline{\lambda}_2^i x_i, G_i x + \underline{\lambda}_2^i x_i - \underline{\lambda}_2^i\} \geq 0, \quad (4.18)$$

$$x \in X \subseteq \{0,1\}^n \quad (4.19)$$

Linearizing (4.17) - (4.19) and introducing necessary linear transformation, we immediately obtain the following new linearization formulation:

$$\overline{NBP}: \min \quad c^T x + e^T s + \underline{\gamma}_1^T x \quad (4.20)$$

$$s.t. \quad (Q + \underline{\Gamma}_2 + \underline{\Gamma}_1) x = y + s + \underline{\Gamma}_2 e \quad (4.21)$$

$$y \leq [\overline{\Gamma}_1 - \underline{\Gamma}_2](e - x) \quad (4.22)$$

$$s \geq 0 \quad (4.23)$$

$$h^T x + e^T z + \underline{\lambda}_1^T x \geq g \quad (4.24)$$

$$(G + \underline{\Lambda}_2 + \underline{\Lambda}_1) x \geq z + \underline{\lambda}_2 \quad (4.25)$$

$$z \leq [\overline{\Lambda}_2 + \underline{\Lambda}_1] x \quad (4.26)$$

$$x \in X \quad (4.27)$$

The following result is trivial to verify.

**Proposition 4.2.**

(a) Problems $\overline{NBP}$ and P are equivalent in the sense that for each feasible solution to one problem, there exists a feasible solution to the other problem having the same objective value.

(b) Let x be part of an optimal solution to Problem $\overline{NBP}$. Then x solves Problem P.

Though BP, $\overline{BP}$ and $\overline{NBP}$ are equivalent, the continuous relaxation of $\overline{NBP}$ can give tighter lower bound due to the trivial fact which follows from the definitions (2.1), (2.2), (4.1)-(4.4) and (4.9) - (4.12):

**Proposition 4.3.**

$$\bar{\gamma}_{1/2} \leq \gamma \max, \quad (4.28)$$

$$\underline{\gamma}_{1/2} \leq \gamma \min, \quad (4.29)$$

$$\bar{\lambda}_{1/2} \leq \lambda \max, \quad (4.30)$$

$$\underline{\lambda}_{1/2} \leq \lambda \min. \quad (4.31)$$

Now we present the strategy to improve the additional inequalities (2.31) which strengthen BP. According to Lemma 4.1, we have

$$x_i(G_i - Q_i)x = \min\{\bar{w}_2^i x_i, (G_i - Q_i)x + \underline{w}_2^i x_i - \underline{w}_2^i\}, \forall i, \quad (4.32)$$

$$x_i(G_i - Q_i)x = \max\{\underline{w}_1^i x_i, (G_i - Q_i)x + \bar{w}_1^i x_i - \bar{w}_1^i\}, \forall i, \quad (4.33)$$

Where

$$\bar{w}_1^i = \max\{(G_i - Q_i)x : x \in \bar{X}, x_i = 0\}, \forall i, \quad (4.34)$$

$$\underline{w}_1^i = \min\{(G_i - Q_i)x : x \in \bar{X}, x_i = 1\}, \forall i, \quad (4.35)$$

$$\bar{w}_2^i = \max\{(G_i - Q_i)x : x \in \bar{X}, x_i = 1\}, \forall i, \quad (4.36)$$

$$\underline{w}_2^i = \min\{(G_i - Q_i)x : x \in \bar{X}, x_i = 0\}, \forall i, \quad (4.37)$$

Applying the linear transformation (2.22) to (4.32) and (4.33), we obtain

$$(\underline{\lambda}_1^i - \gamma_1^i - \bar{w}_2^i)x_i - s_i - z_i \leq 0, \forall i, \quad (4.38)$$

$$(\underline{\lambda}_1^i - \gamma_1^i - \underline{w}_2^i)x_i - s_i - z_i \leq (G_i - Q_i)x - \underline{w}_2^i, \forall i, \quad (4.39)$$

$$(\underline{\lambda}_1^i - \gamma_1^i - \underline{w}_1^i)x_i - s_i - z_i \leq 0, \forall i, \quad (4.40)$$

$$(\underline{\lambda}_1^i - \gamma_1^i - \bar{w}_1^i)x_i - s_i - z_i \leq (G_i - Q_i)x - \bar{w}_1^i, \forall i, \quad (4.41)$$

**Proposition 4.4.** For each i, (4.38) implies (2.31).

*Proof.* Note that (2.31) means $x_i(Q_i x - G_i x - w_{\max}^i) \leq 0$ while (4.38) is $x_i(Q_i x - G_i x - \bar{w}_2^i) \leq 0$

The proof is completed since $\bar{w}_2^i \leq w_{\max}^i$ from their definitions. □

Finally, we point out that the approach adding inequalities (2.31) or our (4.38) - (4.41) to strengthen *BP* is actually a special case of re-linearizing the quadratic part of the Lagrangian function, $\sum_{i=1}^n x_i(Q_i - \theta G_i)x$, where $\theta > 0$ is the Lagrangian multiplier.

Define

$$\bar{w}_1^i(\theta) = \max\{(-Q_i + \theta G_i)x : x \in \bar{X}, x_i = 0\}, \forall i, \quad (4.42)$$

$$\underline{w}_1^i(\theta) = \min\{(-Q_i + \theta G_i)x : x \in \bar{X}, x_i = 1\}, \forall i, \quad (4.43)$$

$$\bar{w}_2^i(\theta) = \max\{(-Q_i + \theta G_i)x : x \in \bar{X}, x_i = 1\}, \forall i, \quad (4.44)$$

$$\underline{w}_2^i(\theta) = \min\{(-Q_i + \theta G_i)x : x \in \bar{X}, x_i = 0\}, \forall i, \quad (4.45)$$

Similarly, we have the following cuts

$$(\theta\underline{\lambda}_1^i - \gamma_1^i - \bar{w}_2^i)x_i - s_i + \theta z_i \leq 0, \forall i, \quad (4.46)$$

$$(\theta\underline{\lambda}_1^i - \gamma_1^i - \underline{w}_2^i)x_i - s_i + \theta z_i \leq (G_i - Q_i)x - \underline{w}_2^i, \forall i, \quad (4.47)$$

$$(\theta\underline{\lambda}_1^i - \gamma_1^i - \underline{w}_1^i)x_i - s_i + \theta z_i \geq 0, \forall i, \quad (4.48)$$

$$(\theta\underline{\lambda}_1^i - \gamma_1^i - \bar{w}_1^i)x_i - s_i + \theta z_i \geq (G_i - Q_i)x - \bar{w}_1^i, \forall i, \quad (4.49)$$

where $\theta > 0$. If $\theta = 1$, they reduce to (4.38) - (4.41).

A good choice may be

$$\theta = \arg\min_{\theta \geq \varepsilon}\|\bar{w}_1(\theta) - \underline{w}_1(\theta)\| \text{ or } (\min_{\theta \geq \varepsilon}\|\bar{w}_2(\theta) - \underline{w}_2(\theta)\|), (4.50)$$

where $\|.\|$ is a norm and $\theta > \varepsilon$ is a fixed real number. Approximately, we take

$$\theta = \arg\min_{\theta \geq 1}\|Q - \theta G\|_F = \begin{cases} \dfrac{trace(QG^T)}{trace(GG^T)}, & \text{if } trace(QG^T) > 0 \text{ and } \|G\|_F \neq 0 \\ 1 & \text{otherwise.} \end{cases} \quad (4.51)$$

where

$$\|G\|_F = (trace(GG^T))^{\frac{1}{2}} \text{ and } trace\, G = \sum_{i=1}^n G_{ii}.$$

## 5. Conclusions

In this article, we discuss small linearizations for the zero-one quadratic minimization problem. We present the equivalence of quadratic terms and piece-wise linear terms for zero-one variables. There are two piece-wise formulations, convex and concave cases. We show the smaller linearization is based on the convex piece-wise objective function and constraints. Linearization generated by our approach is smaller than that in [14]. Our approach can be easily extended to linearize polynomial zero-one minimization problems.

Further tight strategies are also discussed such as enhancing the lower and upper bounding parameters, strengthening existing cuts and adding new cuts.